\documentclass{aa}
\usepackage{graphicx}
\begin{document}
\title{Polarization Studies of Comet C/2000 WM1 (LINEAR)}


   \author{U.C. Joshi, K.S. Baliyan, S. Ganesh}

   \offprints{U.C. Joshi (joshi@prl.ernet.in)}

\institute{Physical Research Laboratory, Ahmedabad 380009, India}
\date{}
\titlerunning{Comet C/2000 WM1 - Polarization study}
\authorrunning{U.C. Joshi et al.}
\date{Received 25 June 2002 / Accepted 23 April 2003
}
\abstract{
Linear polarization observations were carried out on comet C/2000 WM1
with the 1.2m telescope at Mt. Abu Observatory during
November 2001 and March 2002.
The observations in November were at low phase angle ($<~22\degr$) when
the polarization is
negative and where the data for most of the comets are rather meager.
The observations during March were made when the phase angle was
$\sim$ 47\degr. Observations  were
conducted through the IHW narrow band and BVR broad band filters.
Based on these polarization observations we infer that the comet C/2000 WM1
belongs to high polarization class i.e. the dusty comet family.
}

\maketitle

\section{Introduction}
The polarized radiation received from cometary grains contains vital
information which has proved to be very useful in our
understanding of the origin of solar system.
There are two main mechanisms which contribute to the cometary
polarized radiation: (i) sunlight scattered by the cometary
dust particles, and (ii) fluorescence
emission by the cometary molecules. Linear and circular
polarization measurements have been made by several
investigators in the past for many comets.  Most of these
studies are aimed at understanding the nature of polarization
which occurs due to the scattering of the sunlight by cometary
dust particles. The first major efforts for detailed polarization
observations were made for the Comet P/Halley by many groups  (Bastien et
al. \cite{Bastien1986},
Brooke et al. \cite{Brooke1987}, Dollfus \& Suchail \cite{Dollfus1987},
Kikuchi et al. \cite{Kikuchi1987},
Lamy et al. \cite{Lamy1987},
Le Borgne et al. \cite{Leborgne1987}, Metz \& Haefner \cite{Metz},
Sen et al. \cite{Sen1988}).
There were other bright comets after P/Halley such as
Austin, Hyakutake, and Hale-Bopp and polarization observations were reported
for these comets by several groups (e.g. Sen et al. \cite{Sen1991}, Eaton et al. \cite{Eaton1992},
Joshi et al. \cite{Joshi1997}, Ganesh et al. \cite{Ganesh1998}, Furusho et al. \cite{Furusho1999}, Manset \& Bastien \cite{Manset2000}).

A study of  the variation of polarization with phase angle
($\alpha$), defined as the sun-comet-earth angle, and the
wavelength dependence of polarization allows inferences to be drawn about
the size distribution and composition of the scatterers.
Based on the data obtained by various researchers, Dollfus et al. (1988)
established a phase curve describing the variation of the degree
of linear polarization with the phase angle. The study of the dust grains
in comets has been an active area of
investigation for quite some time but the exact nature and
composition of the cometary grains are still not well understood.
Dollfus (\cite{Dollfus1989}) pointed out the possibility of
the grains giving rise to the polarization being large, rough and
dark, resembling fluffy aggregates such as Brownlee particles.The space
mission to Comet Halley made some in-situ measurements and have
contributed new information on the nature of grains in that
comet (Kissel et al. \cite{Kissel1986},  Mazets et al. \cite{Mazets1986},
Levasseur-Regourd et al. \cite{LR1986}).
However, ground based observations
have indicated that the detailed behaviour of grains
differs in different comets. Until more in-situ measurements are carried
out on other comets, the information about dust grains in the comets
has to come mainly from the ground based observations in
conjunction with theoretical models (Krisnaswamy \cite{Krisnaswamy1986},
Xing \& Hanner \cite{Xing1997}, Jockers \cite{Jockers1999a},
Petrova et al. \cite{Petrova2001a},  Petrova et al. \cite{Petrova2001b}).

On the basis of the polarization behaviour of 13 comets,
Chernova(\cite{Chernova1993}) pointed out
the existence of two types of comets: gassy and dusty.
Subsequently Levasseur-Regourd et al. (\cite{LR1996}) and
Hadamcik et al. (\cite{Hadamcik1999})
expanded this work by studying the polarization phase curve for a large
number of comets. It is seen that all the polarization phase curves show
similar behaviour at low phase angles i.e. small negative
values of polarization for phase angle $\alpha$ $<$ 22\degr, increasing nearly linearly in the
range 30 $<$ $\alpha$ $<$ 70\degr \, and reaching a maximum value of 10-30$\%$ in the
phase angle range 90-110\degr. However, for $\alpha$ $>$ 30\degr \, comets
follow two distinct distributions in the polarization-phase angle diagram which led
Levasseur-Regourd et al. (\cite{LR1996}) to propose two classes of comets:
low polarization and high
polarization comets. It is clear now that high polarization comets are dusty
while low polarization comets are gassy.
The maximum value of polarization reaches
about 15\% and 25\%  for the gassy and the dusty comet classes respectively. In fact
the degree of polarization is found higher for
Hale-Bopp than for any other comet previously observed so much so that
Hadamcik et al.(\cite{Hadamcik2002})
proposed to add one more class (Hale-Bopp class) to the classification of comets.

Initially the detection of negative polarization at low phase angle ($\alpha$ $<$ 22$^o$)
in comets was received with a surprise as the negative polarization
was considered more associated with atmosphere-less solar system
objects having a surface made of fluffy layers of small grains, such as Mercury,
the Moon, the asteroids (Dollfus \& Auriere \cite{Dollfus1974},
Gehrels et al. \cite{Gehrels1987},
Zellner \& Gradie \cite{Zellner1976},  Dollfus \& Zellner \cite{Dollfus1979}).
The objects quite different
in nature - dark and bright satellites of the planets, Mercury, asteroids and
cometary comae demonstrate similar negative polarization below phase angle 22\degr.
This phenomenon is observed in a wide spectral range from UV to IR and
the polarization appears to be almost wavelength independent.
Also the cross over angle ($\alpha_{inv}$), slope at the cross over point and polarization
minimum $(P_{min})$ do not show significant wavelength dependence
in most of the cases. However, some faint comets, like 29P/Schwassmann-Wachmann
(Kiselev \& Chernova \cite{Kiselev1979})  or
47P/Ashbrook-Jackson(Kiselev \& Chernova \cite{Kiselev1981}) are reported to
have shown significant variation of polarization with wavelength for $\alpha$ $<$ 25\degr.
These day-to-day variations detected  at a given
phase angle were thought to be related to the activity in comets.
Though the high precision polarimetric
observations of comets covering a large phase angle range
are important to fully characterize the grains,
such data are lacking for the negative branch of the polarization phase curve.
High S/N observations at low phase angles are relatively difficult
due to the comets being faint, as in general comets are
at large distances from the Sun when $\alpha$ is low.
The complete information on the polarization phase curve for a large number of comets
would  be helpful to understand their formation and evolution
and their relation to other solar system bodies.

The apparition of comet C/2000 WM1 provided a good opportunity to
make polarimetric observations at low phase
angles.
During its pre-perihelion approach the comet was conveniently located in the sky
 and was bright enough to achieve a reasonably good S/N ratio at low phase angles.
On 22.67 January 2002 the comet was at a perihelion distance of
0.55AU.
We carried out linear polarization
observations on this comet during November 23-26, 2001 when the phase
angle, $\alpha$, was $<$ 22\degr \, and later in March when $\alpha$ was
$\approx$ 47\degr. The preliminary results from these observations were
reported by Joshi et al (\cite{Joshi2002}). In this communication, the
 results are discussed in detail and compared with the results for other
comets.

\section{Observations and analysis}
Photopolarimetric observations of comet C/2000 WM1 were made with a two channel
photopolarimeter (Deshpande et al. \cite{Deshpande},  Joshi et al. \cite{Joshi1987})
 mounted on the 1.2m telescope of Mt. Abu Observatory operated by Physical
Research Laboratory, Ahmedabad(PRL).  The PRL photopolarimeter
 modulates the polarized component
of the incident light at 41.67Hz with a rotating
super-achromatic half-wave plate in front of a Wollaston prism.
It is equipped with IAU/IHW filters
(see Osborn et al. \cite{Osborn1990})  and
BVR broad band filters. The IAU/IHW filters were acquired about a decade ago
for observations of comet Halley. Since then these were used for
observations of several other comets(Ganesh et al. \cite{Ganesh1998}, Sen et al. \cite{Sen1991},
Joshi et al. \cite{Joshi1987}). The filters are carefully stored in dry atmospheric conditions
to preserve their transmission characteristics. However, to be sure of
their characteristics, their transmission
curves were obtained in the laboratory and compared with the original
curves supplied by the manufacturer. We found that except for the CO$^+$ filter (4260 \AA),
characteristics for all other filters compare very well with the original transmission
characteristics.
It is to be noted that the observations made with the
same set of filters facilitate comparison with other comets observed
earlier,  hence their continued use is justified.
\begin{table}
\caption{Observation log and comet parameters.  Final column (Ap.) is the linear size of the
aperture projected on the comet. }
\begin{tabular}{lccccr}
\hline
Date \& &	$r$&	$\Delta$&		$Phase$&	Moon-& 	Ap. \\
Time(UT)&  AU&AU&deg&set(UT)&kms\\
\hline
&\\
23/11/2001\\
   20:00 &	1.3458&	0.3758&		14.91&	19:38 	&	 7242 \\
   20:30 &	1.3455&	0.3756&		14.94&	      	&        7238 \\
   \\
24/11/2001\\
   21:00 &	1.3291&	0.3630&   	16.80&	20:30	& 	 6995 \\
   21:30 &	1.3287&	0.3627&		16.84&		&	 6990 \\
   \\
25/11/2001\\
   21:30 &	1.3126&	0.3517&		19.08&	21:20 	&	 6778 \\
   22:00 &	1.3123&	0.3515&		19.13&		&	 6774 \\
   \\
26/11/2001\\
   22:00 &	1.2962&	0.3418&		21.73&	22:12 	&   	 6587 \\
   22:30 &	1.2958&	0.3416&		21.79&		&	 6583 \\
   \\
16/03/2002\\
   23:30 &	1.2401&	1.2375&		47.36&	15:24 	&	23848 \\
\\
17/03/2002\\
   00:00 &	1.2404&	1.2374&		47.35&	16:14 	&	23846 \\
   23:30 &	1.2563&	1.2374&		47.03&		&	23846 \\
\\
18/03/2002\\
   00:00 &	1.2567&	1.2374&		47.03&	17:08 	&	23846 \\
&\\
\hline
\end{tabular}
\label{obstab}
\end{table}

Observations were made during November 23-26, 2001 when phase
angle ranged from 14-22$^o$ and then during March 16-18 ($\alpha \approx $ 47\degr).
We attempted observations through IHW filters
in continuum bands 3650/80, 4845/65 and  6840/90.
However, the comet was too faint to achieve good S/N ratios with these
narrow band filters
even for an integration time as long as 10 min.  Also, during the November
observing run the presence of the Moon in the early part of the night hampered
the polarimetric observations of the comet.
In the presence of the Moon, not only the sky background was bright but also
highly polarized and even a minute variation in the sky
resulted in large error in the measured polarization.
At low phase angle the degree of polarization being low ($P \sim $ 1 to 2\%), \,
 the S/N ratio deteriorates in presence of highly polarized Moon light. Therefore
the polarimetric observations were made after the Moon had set and the sky was dark
albeit  at the cost of available observing time. To achieve a good S/N in
a limited available observing time, we switched over to
observations with the broad band - BVR filters of the Johnson and Morgan system.
 The Sun-comet distance during the November
observing run was about 1.3AU. In general comets start showing activity
at such a distance and therefore observations in broad band filters
 are usually expected to be
contaminated by molecular emission. However,  as discussed in the
next section, the magnitudes in all the bands, normalized with respect
to 4845\AA band, do not show any
variation throughout the observing run in November, indicating that
the molecular emission
remained weak and steady during the observing run and hence the use
 of BVR filters is justified.
To allow for comparison, we made a few observations through the IHW
 filter 4845/65 (cf. Table 2).

\begin{figure}[ht] 
\centering{
\includegraphics[width=\columnwidth]{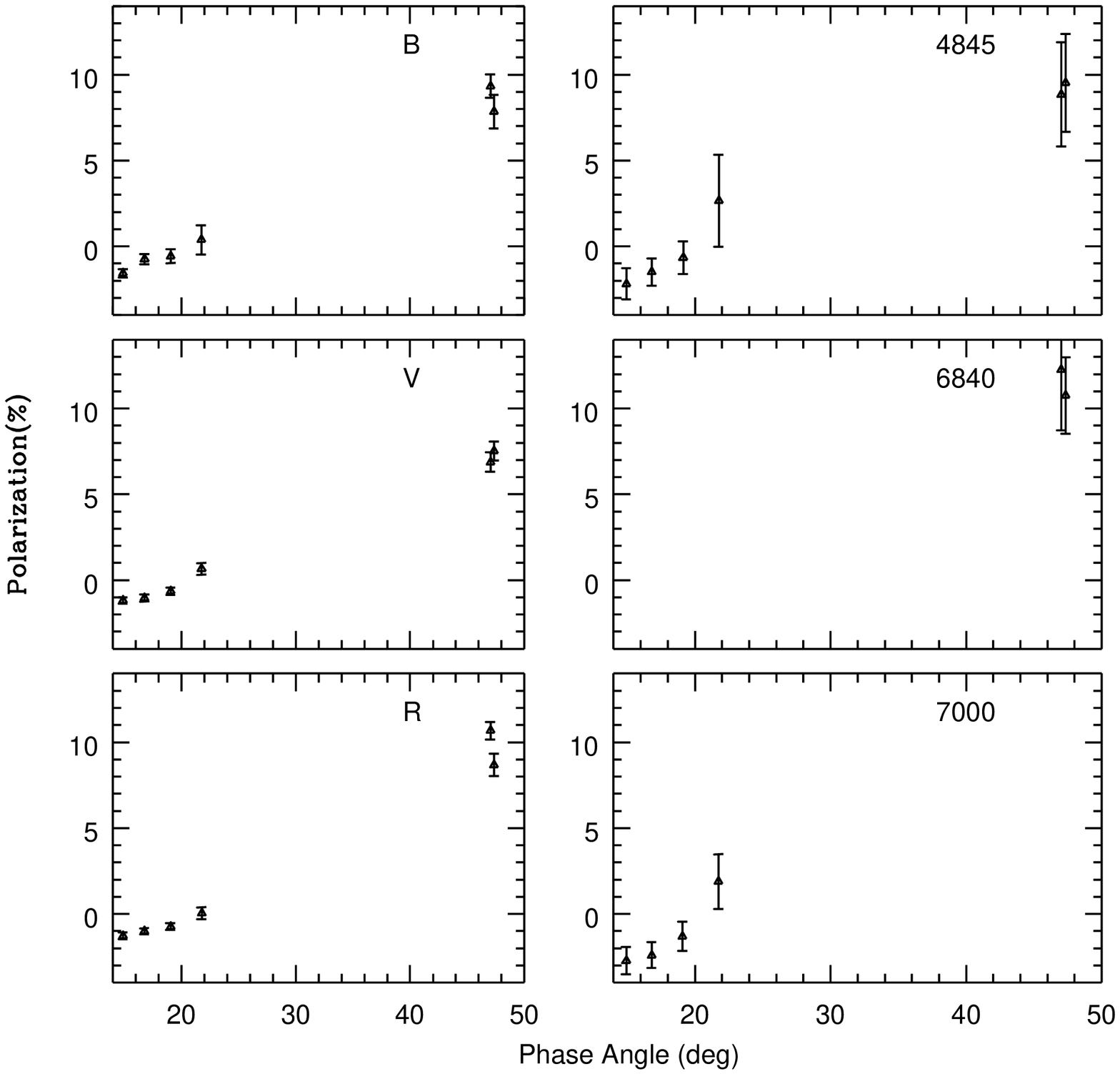}}
\caption {The phase dependence of polarization in
BVR broad bands and IAU/IHW filter bands for the comet C/2000 WM1.}
\label{hb1}
\end{figure}

\begin{figure}[ht] 
\centering{\includegraphics[width=\columnwidth]{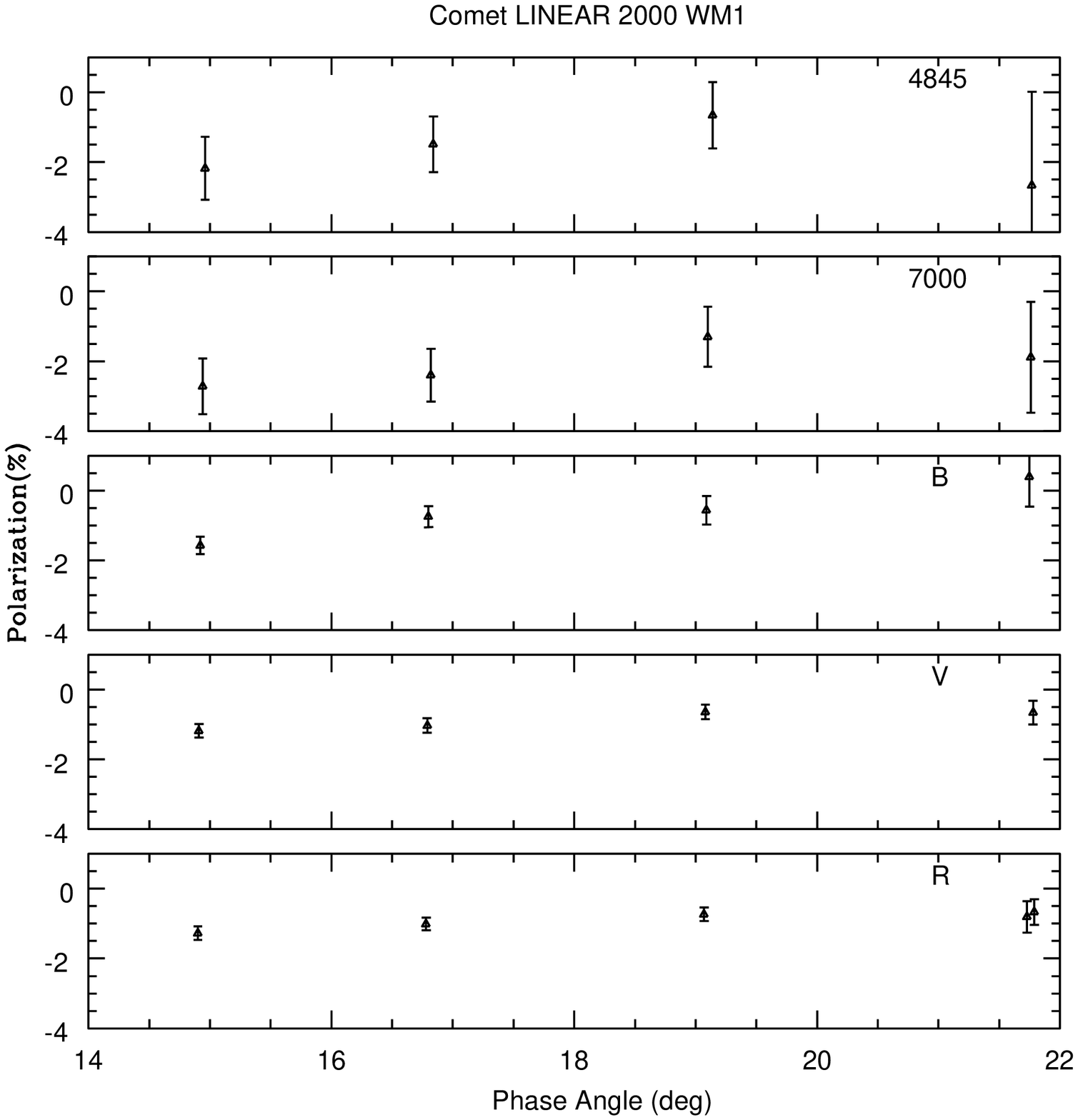}}
\caption{Enlarged version of the phase dependence of polarization of
Comet C/2000 WM1 for lower phase angle}
\label{hb2}
\end{figure}

The observations were taken with an aperture size of 26.5 arcsec centered
on the photocentre of the comet. We took several measurements of shorter integrations (30 to 50
seconds) per filter which were later averaged.  The errors associated with these observations are
estimated as follows.  As described by Joshi et al (\cite{Joshi1987}),  a
least squares fit to the
data provides the degree of polarization and the position angle.  The error in the fit gives the
error in the degree of polarization while the errors in the position angle are obtained using the
equation 8.5.4 given in Serkowski (\cite{Serkowski1974}).
To take care of sky polarization,
observations were made alternately on the photocentre of the comet and
on the region of the sky more than 30 arcmin away from the Sun-comet line.
Polarization standard stars $\phi$ Cas and 9 Gem were observed to calibrate the
observed position angle. Instrumental polarization was much
smaller(0.03\%) than the errors in observations and therefore is neglected.
The journal of the observations along with other informations like phase
angle of the comet, UT, moon set time at Mt. Abu etc is shown in Table 1. Heliocentric
and geocentric distances and the aperture projected on the comet
at the time of observations are also listed. Table 2 lists the
observed values of polarization in BVR bands along with
some observations made through IHW filters.
There are four observing runs in November 2001 (before perihelion passage)
and two in March 2002 (after the perihelion passage).
In addition to the polarization values we also report the normalised magnitudes in
different filter bands. Instrumental magnitudes were first corrected for the atmospheric
extinction and the values thus obtained were normalised with respect to the magnitude
in the continuum band at 4845\AA (Table 2). On 17th March 2002, the observations
were made when the elevation of the comet was low.  Hence, photometric errors being
large, the magnitude values are not considered.
 The polarization phase curves are plotted in
Figure \ref{hb1} and \ref{hb2} and wavelength dependence of polarization is shown in Figure \ref{hb3}.

\begin{table*}
\caption{Polarization observations of comet C/2000 WM1 (LINEAR). Listed entries are
 date and time(UT), total
 integration time(IT) in seconds, filter, degree of polarization(P\%) , error in polarization(Ep\%),
 position angle($\theta$) in equatorial plane, magnitude (normalized with respect to magnitude
 at 4845 band),  and phase angle($\alpha$)
 at the time of observations.  Magnitude values are not listed for 17th March 2002 as the
 photometric errors were larger than 10\%.}
\begin{tabular}{llllllllll}
\hline
Date&	      Time(UT)&  IT &	Filter&   P\%&  ${\epsilon}_p\%$ &   $\theta$& ${\epsilon}_{theta}$ &Mag&  Phase\\
\hline
&\\
23/11/2001&    19 58 50 &  200 &    R  &     1.27&    0.19&   102 & 	4&	-3.24 & 14.90\\
23/11/2001&    20 07 10 &  200&     V &      1.18&    0.19 &  108&      5&	-3.34&  14.91 \\
23/11/2001&    20 13 30 &  200&     B &      1.57&    0.25&   103 & 	4&	-2.80&  14.92 \\
23/11/2001&    20 35 00 &  400&     7000&    2.72&    0.80 &  104 & 	8&	-0.17 & 14.94\\
23/11/2001&    20 46 40 &  450&     4845&    2.18&    0.90 &   110 &      12&    0.0&   14.96\\
&\\
24/11/2001&    20 45 30 &  200 &     R&      1.01&    0.18&    94& 	5&	 -3.22&     16.78\\
24/11/2001&    20 53 30 &  250 &     V&      1.03&    0.21&   102& 	6&	 -3.31&     16.79\\
24/11/2001&    21 04 10  & 450&     B&       0.75&    0.30&    90&     11&	 -2.82&     16.80\\
24/11/2001&    21 16 00  & 500&     7000&    2.40&    0.75&    72& 	9&	 -0.16&     16.82 \\
24/11/2001&    21 30 30  & 600 &    4845&    1.49&    0.80&    93&     15&	  0.0&     16.84 \\
&\\
25/11/2001&    21 26 40&  200&     R&       0.73&    0.19&    67&       7&	 -3.24&     19.07\\
25/11/2001&    21 32 30&  200&     V&       0.64&    0.21&    87&       9&    -3.34&     19.08\\
25/11/2001&    21 38 50&  250&     B&       0.56&    0.41&    84&       20&    -2.95&     19.09\\
25/11/2001&    21 47 40&  450&     7000&    1.30&    0.86&    83&       19&    -0.16&     19.10\\
25/11/2001&    22 08 30&  600&     4845&    0.66&    0.95&    74&       52&     0.0&     19.14 \\
&\\
26/11/2001&    22 11 30&  200&     B&       0.39&    0.85&   177&       52&    -2.87&     21.75\\
26/11/2001&    22 15 50&  150&     7000&    1.9&    1.6&   140&       52&    -0.26&     21.76\\
26/11/2001&    22 21 00&  250&     4845&    2.7&    2.7&   148&       52&     0.0&     21.77\\
26/11/2001&    22 28 00&  200&     V&       0.66&    0.34&   161&       15&    -3.5&     21.78\\
26/11/2001&    22 35 30&  200&     R&       0.04&    0.35&   106&       52&    -3.27&     21.79\\
&\\
16/03/2002&    23 28 53&  180&     R&       8.68&    0.66&    6&        2&    -3.21&     47.36 \\
16/03/2002&    23 34 55&  180&     V&       7.52&    0.55&    1&        2&    -3.55&     47.36 \\
16/03/2002&    23 42 40&  180&     B&       7.85&    0.98&    3&        3&    -2.95&     47.35 \\
16/03/2002&    23 50 30&  360&     6840&   10.8&    2.2&   172&       6&      -0.16&     47.35 \\
16/03/2002&    23 59 59&  300&     4845&    9.5&    2.8&   170&       8&       0.0&     47.35 \\
17/03/2002&    00 11 50&  300&     5140&    9.2&    1.4&    7&        4&      -0.83&     47.35\\
&\\
17/03/2002&    23 47 22&  100&     4845&    13.3&    3.5&    3&        10&   &     47.03 \\
17/03/2002&    23 53 21&  100&     6840&   12.3&    3.5&    19&       8&   &     47.03 \\
18/03/2002&    00 02 51&  100&     B&       9.34&    0.68&    0&        2& &     47.03 \\
18/03/2002&    00 08 00&   60&     V&       6.88&    0.57&    7&        2& &     47.03 \\
18/03/2002&    00 12 22&   60&     R&      10.68&    0.51&    15&       1& &     47.03 \\
18/03/2002&    00 16 31&   60&     5140&    7.9&    1.5&    0&        5&   &     47.03 \\
&\\
\hline
\end{tabular}
\end{table*}

\section{Discussion}
 During the November observing run, the Sun-comet distance changed
only by about 5\% while the Earth-comet distance changed
by about 10\% resulting in a variation of  $\approx$ 20\% in the sampled area on the comet.
This might cause some change in the degree of polarization provided the
sampled region is heterogeneous. The inner coma region of comets
has been found to be quite heterogeneous in the spatial distribution of dust,
which in turn is responsible for the spatial variation of polarization
in the inner region
(Renard \& Hadamcik \cite{Renard1996}, Eaton et al. \cite{Eaton1988},
Eaton et al. \cite{Eaton1991}, Dollfus \& Suchail \cite{Dollfus1987},
Jockers et al. \cite{Jockers1999b}, Furusho et al. \cite{Furusho1999},
Kolokolova et al. \cite{Kolokolova2001}). In the
case of Comet Halley a fair degree of agreement is seen among the
observations made by different groups with different aperture sizes as
long as the polarization is estimated over the whole coma with a large
aperture centered on the nucleus which averages out the effect of
heterogeneity.  In the case of comet Hale Bopp no significant difference
was found in the polarization observed through two
apertures, 26.5 and 52.4 arcsec, corresponding to linear scales of
14318km and 28313km respectively(Ganesh et al. \cite{Ganesh1998}). Similar results
are reported for Hale-Bopp by Manset \& Bastien(\cite{Manset2000}).
In the present case the radius of the area sampled on the coma is relatively large
(6500km to 7200km in diameter, much larger than the inner coma) and
the inhomogeneities in the coma are expected to  be averaged out. The observed
data (Table 2) support this view. We also note that the overall polarization
characteristics of the comet WM1 are similar to comet Halley and Hale-Bopp as is
discussed later. Therefore, the comparison of the polarimetric observations
on different dates for comet C/2000 WM1  is meaningful.

\subsection{Polarization phase curve}
The polarization behavior of the comet C/2000 WM1 is displayed in
Figure \ref{hb1} which shows the degree of polarization for different
spectral bands (BVR and IHW filters: 4845 and 6840) as a function
of phase angle in the range 14-47\degr. There is a good coverage
of phase angle between 14 and 22\degr.
It should be noted that the data are very rare in the
important range of low phase angles. The polarization behavior in this
phase angle range provides clues to the refractory nature of the
grains. Observations were largely made through the broad-band filters to
achieve good S/N ratio. However, some observations were made through
narrow-band IHW filters which also help to  compare with the observations made through
broad-band filters. The normalized magnitudes do not show variation  during the
observing run in November (see Table 2), indicating a weak and steady
 cometary activity. This means
that molecular emission was too weak to have any
significant influence on continuum polarization.  We find from the figure that
the polarization values observed through BVR filters
and the narrow band filters compare very well. In figure \ref{hb2}, we have plotted a magnified view of the
polarization-phase-curve for phase angle $<~22\degr$.  As seen in this figure,
polarization behaviour near $P_{min}$ appears, within the errors of measurement,
to be independent of wavelength.

Near the cross-over phase angle, the degree of polarization is close to zero and hence the position
angle is ill-defined.
Looking at Table 2 we notice that on November 26 the position angle in different
spectral bands deviates from perpendicular
or parallel to the scattering plane when the phase angle is near 22\degr.  The errors associated
with the position angle measurements are large.  Therefore no meaningful conclusions can be drawn
from the position angle.

\subsection{Wavelength dependence of polarization}
\begin{figure}[ht] 
\centerline{\includegraphics[width=\columnwidth]{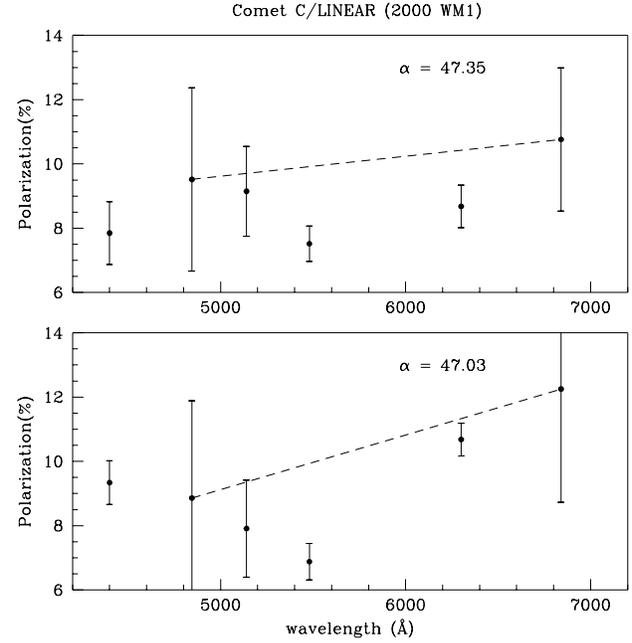}}
\caption{ Wavelength dependence of the degree of polarization for the
comet C/2000 WM1. The polarization values at the continuum bands 4845 and 6840\AA \~are
joined by the dashed line.}
\label{hb3}
\end{figure}

Figure \ref{hb3} displays degree of polarization against the mean wavelength for the
comet when its phase is about 47\degr.
The IHW and BVR filters are indicated in the figure. The figure
 shows that the degree of polarization increases with the wavelength
though the errors are large for IHW filters. In general the values of polarization for
broad band filters are lower compared to the IHW filters as one would expect.
On March 17/18, 2002 when the comet phase was 47.03\degr, the polarization values indicate
an increase over the values of the previous night when the phase was at 47.35\degr. The
observations also indicate a tendency for the polarization colour to be
redder compared to the previous
night which might be due to the increased comet activity  releasing more
small grains. However, since the errors are relatively large, the above
statement  is to be taken with caution.

We notice that the  polarization
observed through the V-band on March 17/18 is lower than what was observed on the
previous night, though the polarization in the R band shows an increase.
The reason is that the R band polarization is less influenced by molecular
 emission compared to the V band.

\subsection{Classification of WM1 based on polarization data}
Based on the extent of polarization at various phase angles, comets are classified in two classes,
high polarization and low polarization class (Levasseur-Regourd et al. \cite{LR1996},
Hadamcik et al. \cite{Hadamcik1999}). We compare the
polarization values for comet C/2000 WM1 in Table 2 with the values from
combined polarization phase curve as given by Levasseur-Regourd et al.(\cite{LR1996}).
 At the 47\degr \, phase angle  the expected
value of polarization in the case of the high polarization class (dusty comets)
 is $\sim$ 11$\%$ whereas for the low polarization
class the expected value is $\sim$ 5.5\%. The observed polarization value
for comet C/2000 WM1 at this phase angle is around 10\%.
The polarization values as observed on March 16 and 17 differ by a small
amount which cannot be due only to the very small change in phase angle but the main
contribution could be from increased cometary activity.
The observed values of the polarization on both the
above dates are close to the mean value of 11\%  as obtained from the common phase curve for the
dusty comets. Therefore the comet C/2000 WM1 belongs to the high polarization class
i.e. to the class of dusty comets.

\subsection{On the nature of the grains}
Figure \ref{hb2} shows a magnified version of P vs $\alpha$ curve for $\alpha$  $<$ 22\degr.
The polarization phase curve of comet C/2000 WM1(Figure \ref{hb1} and \ref{hb2}) is very similar to other bright
dusty comets like comet P/Halley and Hale-Bopp, indicating that the dust particles  have similar
characteristics in all these comets. In the present study the polarization phase curve is
well covered for low phase angle where the polarization is negative. The negative
branch of the polarization phase curve in comets is still not well understood as similar
behaviour is observed in other  atmosphere-less solar system bodies such as Mercury,
the Moon and asteroids (Dollfus \& Auriere \cite{Dollfus1974}, Gehrels et al. \cite{Gehrels1987},
Zellner \& Gradie \cite{Zellner1976}, Dollfus \& Zellner \cite{Dollfus1979}).
The similarity in the polarization behaviour
of the cometary grains and these types of scattering surfaces was noted by several researchers
(Kiselev \& Chernova \cite{Kiselev1981}, Myer \cite{Myer1985}, Bastien et al. \cite{Bastien1986},
Steigmann \& Dodsworth \cite{Steigmann1987}).  It should be noted that while the
cometary coma is optically thin, atmosphere-less solar system bodies are regolith
surfaces and are treated as  optically thick systems.  Therefore, the
objects, which are quite different in nature
(e.g. asteroids, planet Mercury, cometary coma) but show similar negative branch of
polarization, supposedly have an  aggregate structure of dust grains
(Brownlee \cite{Brownlee1985a}, Brownlee\cite{Brownlee1985b},
 West \& Smith \cite{West1991}, Petrova et al. \cite{Petrova2001a},
Petrova et al. \cite{Petrova2001b}).
From the above facts one can infer that negative polarization of light at
small phase angle is a collective effect of light scattering from particles
and not a manifestation of their individual physical properties. Shkuratov and his
colleagues showed that the negative branch of polarization appears if the particles of
the scattering surface or their structural irregularities are comparable to the
wavelength of radiation (Shkuratov \cite{Shkuratov1994},
Shkuratov et al. \cite{Shkuratovetal1994}).
Theoretical modeling of the properties of aggregate
particles confirms that the collective effect of monomers significantly changes the single
scattering characteristics of clusters as compared to those of the independent
particles, and negative polarization can be produced with parameters (aggregate size,
monomer size, wavelength, refractive index) close to those for the particles observed in
cometary coma (West \& Smith \cite{West1991}, Xing \& Hanner \cite{Xing1997},
Petrova et al. \cite{Petrova2001b}).

The  observations show that the negative polarization is almost independent of the
wavelength, which means that dust particle characteristics, i.e. the mean size
and composition, do not play an important role in generating the negative
branch of the phase curve. However, careful examination of the polarization
data of several comets show some minor difference in polarization behaviour
at smaller phase angle as is seen in the Table 2 of Dollfus(\cite{Dollfus1989}).
For example, comets Ashbrook-Jackson (1977g) and Chernykh(1977e) show deeper polarization
minima compared to the other comets listed in their table.
The crossover angle also differs from
comet to comet, though this variation is small and depends weakly on the
wavelength of observation. In the present case the observations show that the
shorter the wavelength, the smaller is the crossover angle i.e.
$\alpha_{inv}$ is smaller for B-band compared to that for the V- and the R-bands. This indicates
that the grains in this comet consist of a silicate core and organic mantle as discussed below.

A model of single scattering of light is quite appropriate for
comets since the density of the dust in
cometary coma is too low for multiple scattering to take place.
Therefore, the negative  polarization observed in comets
at small $\alpha$ can only be explained by an aggregate structure of dust particles.
An aggregate of random structure composed of monomers with size parameter close to 1.5
and refractive index close to (1.65 + i0.05) displays polarization properties
(Petrova \cite{Petrova2001b}) similar to those observed in comets such
as P/Halley, Hale-Bopp, and the present case  of C/2000 WM1. These properties include
 negative polarization at small $\alpha$,  weak dependence of
inversion angle $\alpha _{inv}$ on wavelength, red colour of cometary dust and
the degree of polarization independent of $\lambda$ at low phase angles. 

Petrova et al.(\cite{Petrova2001b}) have shown that a composite
structure of aggregate particles
resulting in the interaction of monomers in the light scattering process is
responsible for the negative polarization at small phase angles if the monomer size
is comparable to the wavelength of radiation. Comparison of the model calculations of
Petrova et al.(\cite{Petrova2001b}) with the present observations indicates that
the random structure
composed of monomers of size parameter close to 1.5 and the refractive index close to
(1.65 + i0.05) fits the observed polarization curve of WM1. Therefore
the dust grains appear to be a mixture of silicate and organics.
The model also predicts
an increase in the degree of linear polarization  with wavelength at the larger
phase angles. This is what we noticed during the observing run in
March 16-17, 2002 when the phase angle was approximately 47\degr. We also note
that $\alpha _{inv}$ weakly depends on $\lambda$, being smaller
for shorter wavelength.
 If the imaginary part of the refractive index decreases
 with $\lambda$,  the negative
branch of polarization and the inversion angle slightly depend on
$\lambda$ (Petrova et al. \cite{Petrova2001b}),
characteristics of organic material. This study suggests that grains with a
 silicate core and organic mantle represent a realistic model. Fluffy aggregate
 of such monomers can explain the negative polarization and other
characteristics of the observed negative branch of the P vs. $\alpha$ curve.

\section{Conclusions}
This work reported linear polarization observations of
comet C/2000 WM1(LINEAR) for the low
phase angle where the data are rather rare and also at a phase angle  near 47\degr.
Our study based on the polarization observations leads to the following conclusions:
\begin{enumerate}
\item Comet C/2000 WM1 belongs to the high polarization
class i.e. the dusty comet family.
\item The negative branch of polarization is explained to be due to the
scattering of light by the aggregate grains with
monomer size comparable to the wavelength of radiation. The mutual influence of the
monomers composing aggregate
particles produces the negative polarization. The present observations are
explained if the monomer size
parameter is close to 1.5 and refractive index close to $(1.65 + i0.05)$.
Possibly, the dust grains are composed
of a silicate core and organic mantle.
\item The observations on March 17, 2002 indicate enhanced cometary activity.
\end{enumerate}

\begin{acknowledgements}
The work reported here is supported by the Department of Space,
Government of India. We are thankful to
Miss Chhaya R. Shah for the computational assistance.
We express our thanks to the referee,
Dr. P. Bastien, for his very constructive remarks.
\end{acknowledgements}

\end{document}